\documentclass[aps,prl,reprint,groupedaddress]{revtex4-1}

\usepackage{graphicx}
\usepackage{amsmath}

\begin{document}


\title{Interferometric evanescent wave excitation of nano-antenna for ultra-sensitive displacement and phase metrology}


\author{Lei Wei}
\email{lei.wei@kcl.ac.uk}
\affiliation{Department of Physics, King's College London, Strand, London, WC2R 2LS, United Kingdom}
\author{Anatoly V. Zayats}
\affiliation{Department of Physics, King's College London, Strand, London, WC2R 2LS, United Kingdom}
\author{Francisco J. Rodr\'{i}guez-Fortu\~{n}o}
\affiliation{Department of Physics, King's College London, Strand, London, WC2R 2LS, United Kingdom}

\date{\today}

\begin{abstract}
We propose a method for ultra-sensitive displacement and phase metrology based on the interferometric evanescent wave excitation of nano-antennas. We show that with a proper choice of nano-antenna, tiny displacements or relative phase variations can be converted into sensitive scattering direction changes in the Fourier $k$-space. These changes stem from the strong position dependence of the imaginary Poynting vector orientation within interfering evanescent waves. Using strongly-evanescent standing waves, high sensitivity is achieved in the nano-antenna's zero scattering direction, which varies linearly with displacement over a long range. With weakly-evanescent wave interference, even higher sensitivity to tiny displacement or phase changes can be reached around chosen location. The high sensitivity of the proposed method can form the basis for many applications.
\end{abstract}


\maketitle

Sensitive optical metrology of tiny displacements has been an enabling technology for modern science and engineering. Interferometric methods\cite{NBobroff1993} are widely deployed by measuring the relative phase differences introduced by tiny physical displacements. The giant Michelson interferometer of LIGO that enables the detection of gravitational waves demonstrates the power of this technique\cite{BPAbbott2016}. Another technique to measure displacements is based on measuring beam deflections by means of differential detection schemes which offers similar fundamental detection limits\cite{CAJPutman1992,SMBarnett2002}. These techniques form invaluable tools for applications that require sub-nm positioning precisions, like single molecules tracking\cite{JGelles1988,JOrtegaArroyo2015,FBalzarotti2017}, force sensing\cite{GMeyer1988,BSanii2010,AGloppe2014,VBlums2018}, stage stabilisation for localization nanoscopy\cite{LNugentGlandorf2004} as well as overlay metrology for semiconductor industry\cite{AJdenBoef2016}. 

Apart from the evolving instrumentation, further advances in the field have been made by the introduction of new concepts to improve the signal to noise ratio such as squeezed light\cite{NTreps2002,RCPooser2015} and weak value amplification\cite{PBenDixon2009,MDTurner2011}. With the development of beam shaping and nanofabrication, a new position sensing concept that involves the interaction of structured light and structured matter such as optical nano-antennas has also been demonstrated in recent years\cite{MNeugebauer2016,ZXi2016,ZXi2017,ABag2018}. Nano-antennas that support both electric and magnetic modes, like high index dielectric nanoparticles, are of special interest recently\cite{AIKuznetsov2016} as their scattering can be shaped by the full electromagnetic nature of the excitation field\cite{PWozniak2015,TDas2015,ZXiOL2016,LWEI2017}. Even in the simplest case of a dipolar nanoparticle, the induced electric and magnetic dipoles can interfere and result in uni-directional scattering, for instance when Kerker's condition is fulfilled \cite{WLiu2018}. Furthermore, the excitation field can be designed in such a way that this uni-directional scattering condition occurs only at certain locations in the excitation field landscape\cite{MNeugebauer2016,ZXi2016,ZXi2017,ABag2018}. Moving the particle slightly away from these locations will diminish the scattering directionality and the relative displacement can thus be measured with a differential technique. In Ref. \cite{ZXi2016}, the singular points/lines of doughnut or Hermite Gaussian beams are deployed around which the displacement is determined. 
The complex topology of tightly focused vector beams enables Kerker's condition to be met in the transverse focal plane. With a differential detection of the pure scattering pattern at the back focal plane (i.e. the $k$-space), relative displacement down to {\r{A}}ngstr\"om level has been demonstrated\cite{ABag2018}. Standing waves formed by counter-propagating beams in free space have been proposed as excitation fields for a metallic nanowire\cite{ZXi2017}. This method predicts theoretically ultra high relative displacement sensitivity around the nodal point where the electric field is zero but magnetic field is at its maximum. Being of extremely weak magnetic dipole polarizability, the metallic nanowire serves an ideal nano-antenna option to fulfill Kerker's condition in the close vicinity of the nodal point. As the nanowire moves across the point, its uni-directional scattering switches the direction and the closer the condition is reached to the nodal point the higher the sensitivity is. However, this advantage is at the same time its disadvantage: due to working at the node, there is an extremely weak total scattering power which is difficult to detect. In addition, both free space focused beams and standing wave excitations suffer from the fact that the unwanted incident light enters the detection $k$-space. In order to detect the pure directional scattering of the nanoparticle, the k-space influenced by both the incident and scattering light is often cropped\cite{MNeugebauer2016,ABag2018}, with a loss of useful information. Finally, these methods have a limited spatial range because high displacement sensitivity exists only near the region of maximum directionality locations. In this Letter, we propose a technique which overcomes all of these drawbacks. We exploit the unique topology of the imaginary Poynting vector of interfering evanescent waves. By using the interfering evanescent waves as excitation fields for dipolar nano-antennas, the transverse displacement or relative phase difference is translated into angular changes in the scattering momentum space with the whole $k$-space being accessible. We further show that this approach can not only achieve extremely high nonlinear displacement/phase sensitivity within a small range, but also allows achieving relatively high and linear displacement sensitivity over a large dynamic range. 

Different from free propagating waves, evanescent waves exhibit unique properties such as transverse spin angular momentum and intrinsic transverse spin-momentum locking\cite{KYBliokh2015}. A nanoparticle placed inside it acts as an antenna that converts local electromagnetic fields into far-field scattering. The special properties of evanescent waves used as excitation fields enable effective directional control of the nano-antenna's scattering\cite{LWei2018} with applications ranging from tunable displays\cite{JXiang2018} to the generation of interesting optical forces\cite{MNietoVesperinas2010,KYBliokh2014}. Interferometric evanescent waves have been used as structured illumination in Total Internal Reflection Fluorescence (TIRF) microscopy which doubles the imaging resolution and improves the contrast by minimizing the amount of incident light going into the detector\cite{DLi2015}. Its electromagnetic properties are yet to be fully exploited, and combined with designed nano-antennas unleashes its full potential in measuring deeply sub-wavelength displacements and tiny phase changes. 

\begin{figure}[!ht]
\centering
\includegraphics[width=0.42\textwidth]{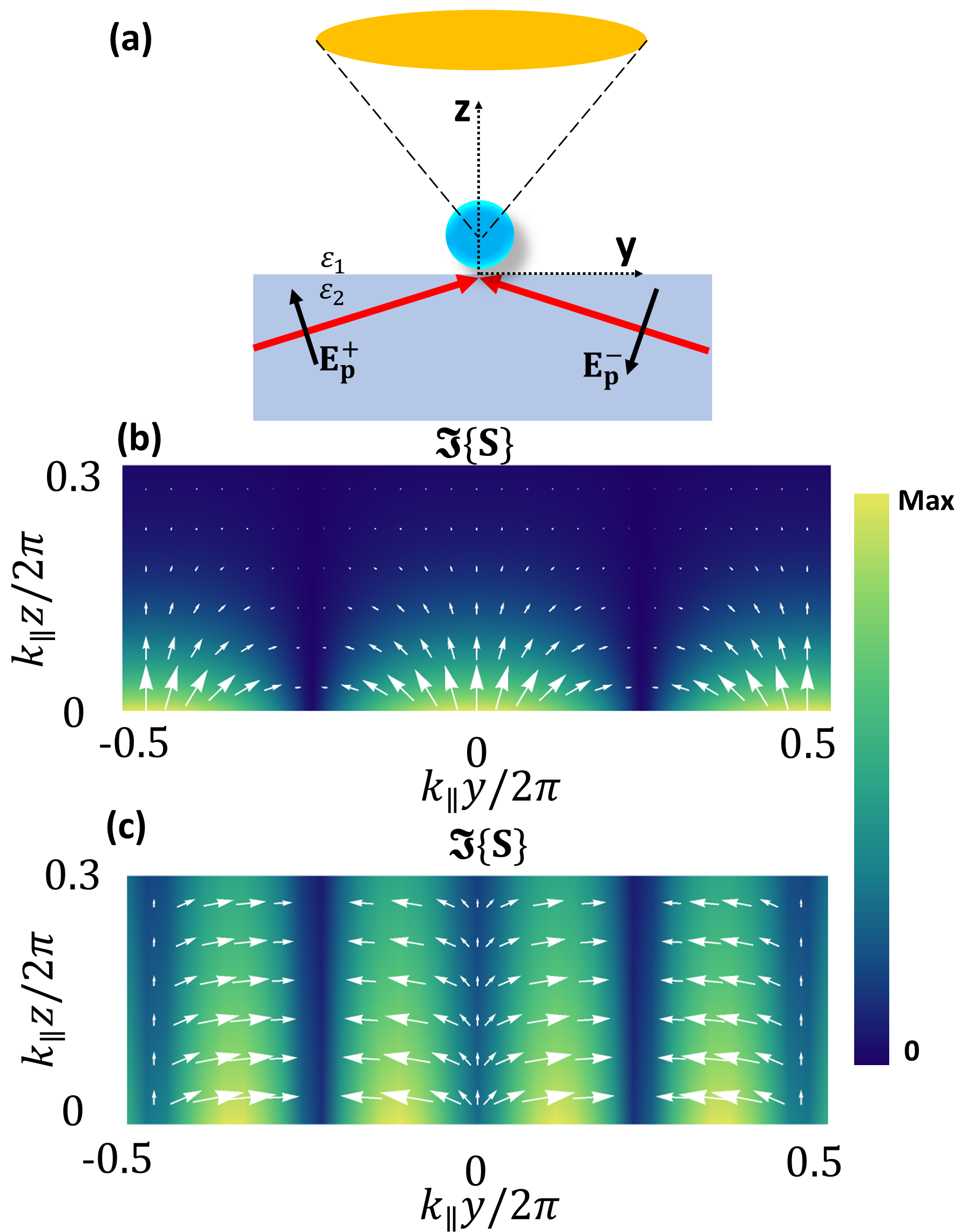}
\caption{(a). A dipolar nanoparticle is excited by two counter-propagating TM polarized evanescent waves. The scattering of the particle is measured at the back focal plane of a lens, i.e. in the $k$-space. (b, c) The imaginary Poynting vector (color represents magnitude and white arrows its orientation) of the standing wave due to the interference of two evanescent waves with opposite transverse wavevectors (b) $\pm k_{\parallel}=\pm 2k_0$, (c) $\pm k_{\parallel}=\pm 1.01k_0$, where the phase difference between the two waves is set to $\Delta\phi=0$.}
\label{fig:1}
\end{figure}
Interfering evanescent standing waves can be set up at an interface, by illuminating with two transverse magnetic (TM or p) polarized incident plane waves $\mathbf{E}_{\mathbf{p}}^+$ and $\mathbf{E}_{\mathbf{p}}^-$  from a higher refractive index substrate above the critical angle, as illustrated in Fig. \ref{fig:1}(a). Consider two incident plane waves with opposite transverse wavevectors $\pm k_{\parallel}$ having exactly the same amplitude but a phase difference $\Delta\phi$. Assuming $\varepsilon_1=1$ and $\varepsilon_2>\varepsilon_1$, the fields of the interfering evanescent wave can be written as:
\begin{align}
\mathbf{E}=&\left[0, -\frac{\gamma_{z}}{k_0}\cos(k_{\parallel}y+\Delta\phi/2), \frac{k_{\parallel}}{k_0}\sin(k_{\parallel}y+\Delta\phi/2)\right]\nonumber\\
&\times2\mathrm{i}E_p\mathrm{exp}(-\gamma_z z),  \nonumber\\ \label{eq:1}
\mathbf{H}=&\left[\frac{1}{Z_0}\cos(k_{\parallel}y+\Delta\phi/2), 0,0\right]\times2E_p\mathrm{exp}(-\gamma_z z),
\end{align}
where $k_0=2\pi/\lambda$ is the wavevector of incoming light, $Z_0$ is the impedance in vacuum and $\gamma_z=\sqrt{k_{\parallel}^2-k_0^2}$ is the imaginary part of the z component of the wave-vector. In this Letter, a time dependence of $e^{-\mathrm{i}\omega t}$ is assumed.

A single TM polarized evanescent wave has a real time-averaged Poynting vector component $S_y$ along the propagation direction and an imaginary Poynting vector component $S_z\ge0$ along the direction of evanescent decay\cite{LWei2018}, while both components are invariant in the $y$ direction. In contrast, the standing wave interference of two TM polarized evanescent waves shown in Eq. \ref{eq:1} results in a purely imaginary time-averaged Poynting vector $\mathbf{S}=\frac{1}{2}(\mathbf{E}\times\mathbf{H}^*)$:
\begin{align}
\mathbf{S}=&\left[0, \frac{k_{\parallel}}{k_0}\sin(2k_{\parallel}y+\Delta\phi), \frac{\gamma_z}{k_0}\left(1+\cos(2k_{\parallel}y+\Delta\phi)\right)\right] \nonumber\\
&\times\mathrm{i}\frac{E_p^2}{Z_0}\mathrm{exp}(-2\gamma_zz). \label{eq:2}
\end{align}
It can be seen that, just as single TM-polarized evanescent fields, the imaginary Poynting vector component of the interference field along the direction of decay is nonnegative $S_z\ge0$. However, its orientation strongly depends on transverse location $y$. At any height $z$, the direction of the imaginary Poynting vector $\Im\{\mathbf{S}\}$ can be determined by: 
\begin{align}
\tan\theta_{\mathrm{s}}=\frac{S_z}{S_y}=\frac{\gamma_z}{k_{\parallel}}\frac{\cos(k_{\parallel}y+\Delta\phi/2)}{\sin(k_{\parallel}y+\Delta\phi/2)}, \label{eq:3}
\end{align}
where $\theta_{\mathrm{s}}$ is the angle between $+\mathbf{\hat{y}}$ and the imaginary Poynting vector $\mathbf{S}$. 
Fig. \ref{fig:1}(b) and (c) show two distinctively different scenarios of Poynting vector's dependencies on the transverse position $y$ inside the standing waves built by the interference of evanescent waves with opposite transverse wavevectors $\pm k_{\parallel}$. Fig. \ref{fig:1}(b) represents the case that $k_{\parallel}/k_0=2$ is relative large while Fig. \ref{fig:1}(c) represents the case with $k_{\parallel}/k_0=1.01$ slightly above 1, corresponding to illumination just above the critical angle. The orientation of $\Im\{S\}$ strongly depends on the $y$-location. A nano-antenna can be used as a local probe of this fast changing Poynting vector to detect the displacement or phase changes by measuring its scattering. The sensitivity of this technique will depend on how fast $\theta_{\mathrm{s}}$ changes with respect to location $y$. As can be derived from Eq. \ref{eq:3}, in the limiting case of illuminating angles well above the critical angle we have $k_{\parallel}/k_0\gg 1$ and $\gamma_z/k_{\parallel}\to 1$: 
\begin{align}
\frac{\mathrm{d}\theta_{\mathrm{s}}}{dy}\approx-k_{\parallel}, \label{eq:4}
\end{align}
the angle $\theta_{\mathrm{s}}$ is changing linearly with the transverse position $y$ across the entire period of the standing wave in Fig. \ref{fig:1}(b). In the limiting case that $k_{\parallel}/k_0$ is slightly above one (and therefore $\gamma_z/k_0$ slightly above zero), the angle of imaginary Poynting vector $\theta_{\mathrm{s}}$ varies in a highly nonlinear manner along $y$ and its rate of change is strongly dependent on the transverse location in Fig. \ref{fig:1}(c). For the fields at a position near $k_{\parallel}y/2\pi=1/4$, one can derive from Eq. \ref{eq:3} that
\begin{align}
\frac{\mathrm{d}\theta_{\mathrm{s}}}{dy}|_{y\to\pi/(2k_{\parallel})}=-\gamma_z, \label{eq:5}
\end{align}
which means the imaginary Poynting vector is changing orientation very slowly around such location. However, for the fields at a position very near $y=0$, one has:
\begin{align}
\frac{\mathrm{d}\theta_{\mathrm{s}}}{dy}|_{y\to0}=-\frac{k^2_{\parallel}}{\gamma_z}, \label{eq:6}
\end{align}
where the imaginary Poynting vector is changing at an extremely high rate around $y=0$ as $\gamma_z\to0$.

As our probe, we consider here a nanoparticle with both electric and magnetic dipole polarizabilities $\alpha_{\mathrm{e}}=\frac{i6\pi\varepsilon_0}{k_0^3}a_1$ and $\alpha_{\mathrm{m}}=\frac{i6\pi}{k_0^3}b_1$, where $\varepsilon_0$ is the vacuum permittivity, $a_1$ and $b_1$ are the Mie coefficients of the electric and magnetic dipole modes. Such nanoparticles can be high index dielectric, core-shell or even metallic nanoparticles as long as their dipole modes dominate in the spectral region of interest \cite{PWozniak2015,TDas2015,ZXiOL2016,LWEI2017,ZXi2017}. For simplicity in explaining the concept of displacement metrology with interferometric evanescent waves, we further neglect the substrate's effect on the particle's scattering. Under such assumptions, the scattering of the nanoparticle is equivalent to the radiation of a source with respectively induced electric and magnetic dipole moments $\mathbf{p}=\alpha_{\mathrm{e}}\mathbf{E}$ and $\mathbf{m}=\alpha_{\mathrm{m}}\mathbf{H}$, where $\mathbf{E}$ and $\mathbf{H}$ are the incident interferometric evanescent fields in Eq. \ref{eq:1} at the center of the nanoparticle. The interference of the radiation fields of the induced electric and magnetic dipoles could result in directional scattering of the nanoparticle under certain conditions. The preferred scattering direction can be shown \cite{MNietoVesperinasOE2010} to be always along the direction of $\Re\{\mathbf{p}\times\mathbf{m^*}\}=\Re\{\alpha_{\mathrm{e}}\alpha_{\mathrm{m}}^*\}\Re\{\mathbf{E}\times\mathbf{H}^*\}-\Im\{\alpha_{\mathrm{e}}\alpha_{\mathrm{m}}^*\}\Im\{\mathbf{E}\times\mathbf{H}^*\}$. For free-space propagating waves, it is purely determined by the particle's dipole polarizabilities $\Re\{\alpha_{\mathrm{e}}\alpha_{\mathrm{m}}^*\}$ and the real Poynting vector. For instance, when $a_1=b_1$ so that the Kerker's condition is fulfilled and $\Re\{\alpha_{\mathrm{e}}\alpha_{\mathrm{m}}^*\}>0$, the particle has zero scattering in the opposite direction of light propagation, i.e. zero backscattering\cite{YHFu2013, WLiu2018}. However, when illuminated by the interfering fields in Eq. \ref{eq:1}, the preferred scattering direction along $-\Im\{\alpha_{\mathrm{e}}\alpha_{\mathrm{m}}^*\}\Im\{\mathbf{E}\times\mathbf{H}^*\}$ is purely determined by the particle's dipole polarizabilities $\Im\{\alpha_{\mathrm{e}}\alpha_{\mathrm{m}}^*\}$  and the imaginary Poynting vector. We shall show in the remaining part of this letter that with a proper choice of nano-antennas, the highly $y$-dependent imaginary Poynting vector in Eq. \ref{eq:4} and Eq. \ref{eq:6} can be converted into scattering direction changes in the Fourier $k$-space, which in return serves as a sensitive means to detect lateral displacements or phases. 

We will first consider a dipolar nanoparticle placed in the interference field shown in Fig. \ref{fig:1}(b) of two TM-polarized evanescent waves with relatively large transverse wavevectors $\pm k_{\parallel}=\pm2k_0$ and a phase difference $\Delta\phi=0$. Consider a particle whose Mie coefficients fulfill $a_1=(-\mathrm{i}k_0/\gamma_z)b_1$, realistic implementations of which were proposed in Ref. \cite{LWei2018}. It can be shown from Eq. \ref{eq:1} that Kerker's condition is satisfied, independent of the position, such that the transverse components of the induced electric and magnetic dipoles always fulfill the relation $p_y=-m_x/c$, with $c$ being the speed of light. As a result, the scattering of the nanoparticle along the $-\hat{\mathbf{z}}$ direction is fixed to be zero no matter where it is located in the standing wave. More generally, the scattering cross section along any direction $\hat{\mathbf{k}}$ can be found proportional to $|\mathbf{p}^*\cdot\tilde{\mathbf{E}}_{\mathbf{k}}+\mathbf{m}^*\cdot\mu_0\tilde{\mathbf{H}}_{\mathbf{k}}|^2$, following Fermi's golden rule which determines the coupling strength of the induced dipoles to the plane wave field vectors $[\tilde{\mathbf{E}}_{\mathbf{k}}, \tilde{\mathbf{H}}_{\mathbf{k}}]$ along wavevector $\mathbf{k}$\cite{MFPicardi2017}. 

\begin{figure}[!ht]
\centering
\includegraphics[width=0.48\textwidth]{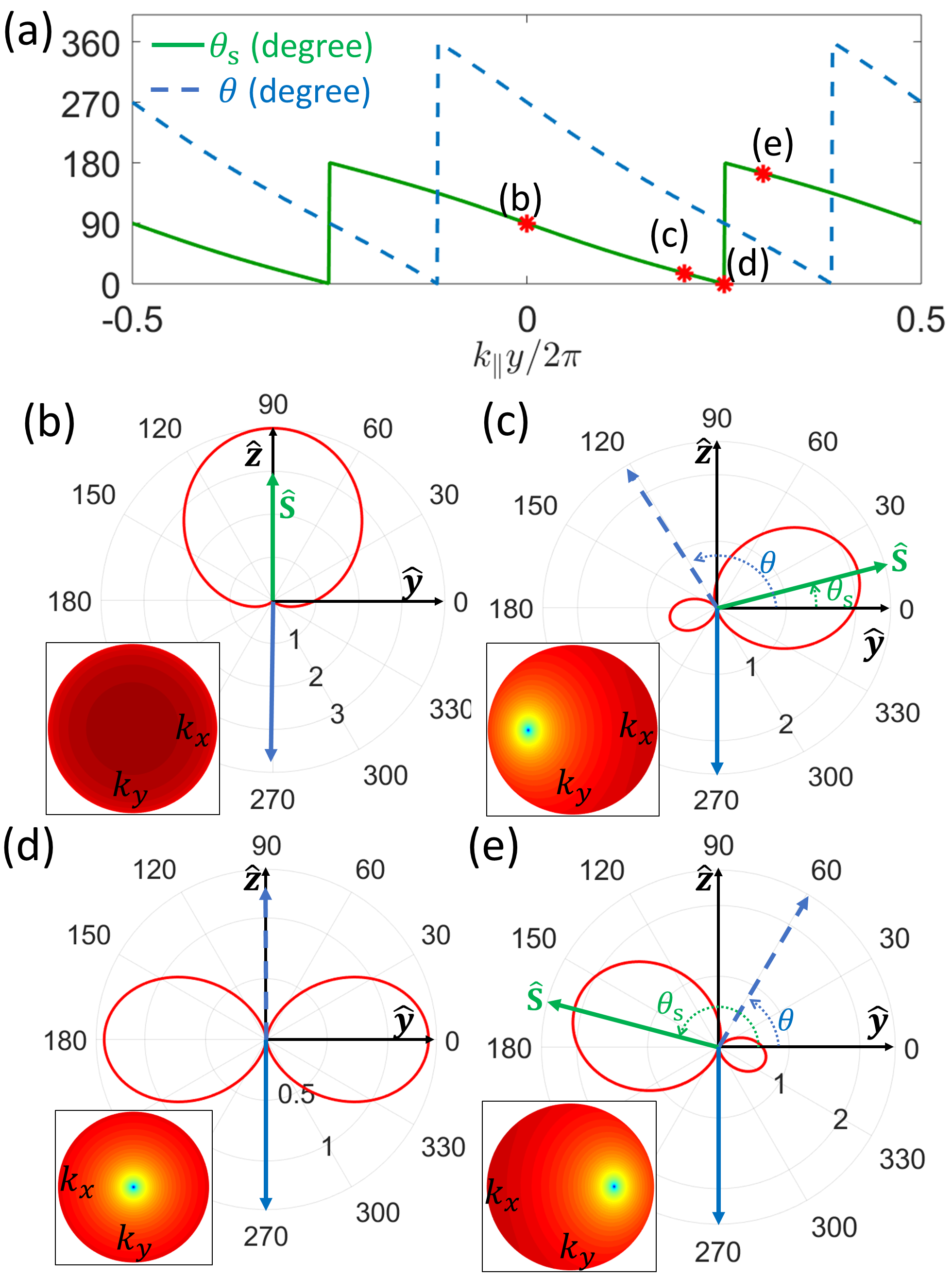}
\caption{(a). $\theta_{\mathrm{s}}$ is the angle from $+\hat{\mathbf{y}}$ to the imaginary Poynting vector of the interferometric evanescent fields with $\pm k_{\parallel}=\pm2k_0$ shown in Fig. \ref{fig:1}(b), while $\theta$ is the angle from $+\hat{\mathbf{y}}$ to the wavevector direction along which the scattering is zero; The scattering patterns of a dipolar particle are shown(the electric and magnetic dipole coefficients fulfill $a_1=\frac{\mathrm{-i}}{\sqrt{3}} b_1$) placed at various locations: (b). $y=0$,  (c). $y=\lambda/4-\lambda/40$, (d). $y=\lambda/4$; (e). $y=\lambda/4+\lambda/40$. The green solid arrows represent the imaginary Poynting vectors, the blue solid arrow represents the fixed zero scattering direction $-\hat{\mathbf{z}}$, the blue dashed arrows represent the other zero scattering direction, while the insets in (b)-(e) show the scattering pattern (logarithm $\mathrm{log}_{10}$ scale, dark red for maximum while deep blue represents zero scattering) in the $k-$space of the Fourier lens above the nanoparticle.}
\label{fig:2}
\end{figure}

In the $k_x=0$ plane, due to the TM nature of the excitation field, the induced dipoles can only couple to the TM-polarized field vectors along $\mathbf{k}=(0, k_y, k_z)=(0,k_0\cos\theta, k_0\sin\theta)$ where $\theta$ is the angle from $+\hat{\mathbf{y}}$ to $\mathbf{k}$ as defined in Fig. \ref{fig:2}, $\tilde{\mathbf{E}}_{\mathrm{TM}, \mathbf{k}}=(0, -\sin\theta, \cos\theta)$ and $\tilde{\mathbf{H}}_{\mathrm{TM}, \mathbf{k}}=(1/Z_0, 0, 0)$. By solving $\mathbf{p}^*\cdot\tilde{\mathbf{E}}_{\mathrm{TM}, \mathbf{k}}+\mathbf{m}^*\cdot\mu_0\tilde{\mathbf{H}}_{\mathrm{TM}, \mathbf{k}}=0$, we can explicitly find two directions of zero scattering: one is fixed at $-\hat{\mathbf{z}}$ which is independent of $y$ and the other directly related to the direction of the imaginary Poynting vector:
\begin{align}
\theta=2\theta_{\mathrm{s}}+\frac{2m+1}{2}\pi,\,\,\,m\,\,\mathrm{is\,\,integer} \label{eq:7}
\end{align}
which, just as the imaginary Poynting vector, is highly dependent on the transverse location $y$. It is easy to observe from Eq. \ref{eq:7} that this zero scattering direction doubles the sensitivity of the imaginary Poynting vector to displacement changes ($\frac{\mathrm{d}\theta}{\mathrm{d}y}=2\frac{\mathrm{d}\theta_{\mathrm{s}}}{\mathrm{d}y}$). This can also be understood by the fact that these two zero scattering directions are symmetric to the axis of the imaginary Poynting vector, so by locking one direction to $-\hat{\mathbf{z}}$, the other one rotates at twice the rotation angle of the symmetry axis, i.e. the imaginary Poynting vector.

Fig. \ref{fig:2} demonstrates how this concept applies to displacement metrology with Fig. \ref{fig:2}(a) showing the $y$-dependence of the imaginary Poynting vector angle $\theta_{\mathrm{s}}$ and the zero scattering direction angle $\theta$, while Fig. \ref{fig:2}(b)-(e) show the scattering patterns of the nano-antenna at various transverse locations of the interference field. At location $y=0$, the scattering pattern in Fig. \ref{fig:2}(b) shows the only zero scattering direction along $-\hat{\mathbf{z}}$ where one sees the maximum scattering direction along the direction of the imaginary Poynting vector at the $k-$space of a lens above the nanoparticle. More interestingly, in the region $\pi/4<k_{\parallel}y<3\pi/4$, one observes the zero scattering direction rapidly moving with $y$ in the upper half space through the Fourier space of the lens. The scattering appears as the one of a vertical electric dipole at the location $y=\lambda/4$, with zero scattering along $+\hat{\mathbf{z}}$. With displacements of $\Delta y=\pm\lambda/40$ around $y=\lambda/4$, as shown respectively in Fig. \ref{fig:2}(c) and (e), changes of zero scattering angle $\Delta\theta=\mp31.43^{\circ}$ are introduced in the Fourier $k$-space. This corresponds to a sensitivity $\Delta\theta/\Delta y$ about $-1257.2^{\circ}/\lambda$, which means a change of $1^{\circ}$ in the zero scattering direction can resolve a displacement of $\lambda/1257.2$. Using even larger $k_{\parallel}/k_0$, the sensitivity can reach up to $-(\frac{k_{\parallel}}{k_0})720^{\circ}/\lambda$. Additionally, both $\theta_{\mathrm{s}}$ and $\theta$ exhibit nearly linear relations with $y$ in Fig. \ref{fig:2}(a) as expected from Eq. \ref{eq:4}. This linear relation with $y$ would enable the detection of deeply sub-wavelength displacements as well as its displacement direction over long ranges.

\begin{figure}[!ht]
\centering
\includegraphics[width=0.48\textwidth]{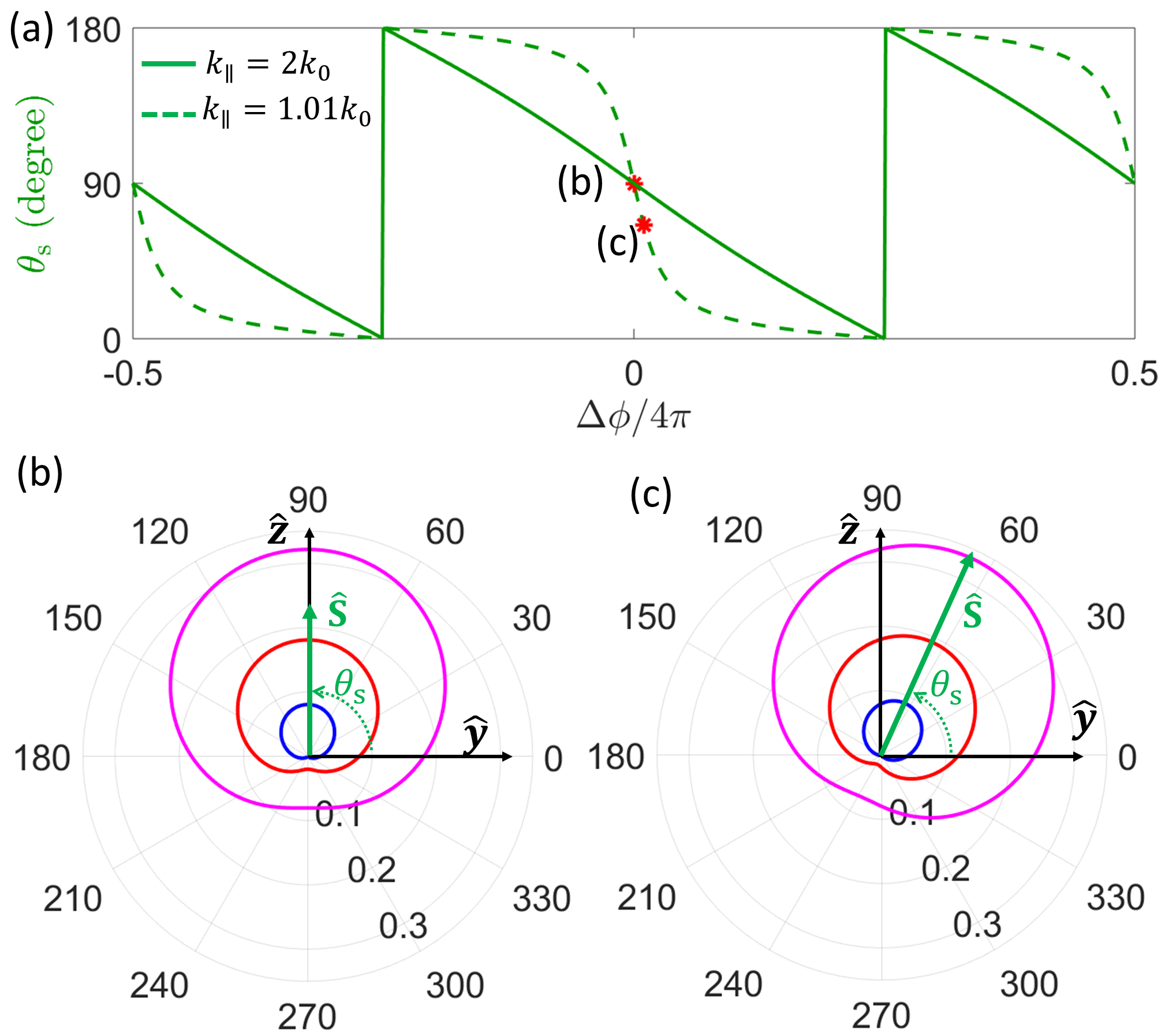}
\caption{(a). The dependence of the imaginary Poynting vector angle $\theta_{\mathrm{s}}$ on the relative phase difference $\Delta\phi$ at a fixed location $y=0$ of the interferometric evanescent field with $\pm k_{\parallel}=\pm2k_0$ and $\pm k_{\parallel}=\pm1.01k_0$ respectively; The scattering patterns are shown for a dipolar particle placed at $y=0$ of the interfering fields with different relative phases: (b). $\Delta\phi=0$,  (c). $\Delta\phi=\pi/25$. The radiation patterns of colour blue, red and magenta in (b) and (c) depict nanoparticles with fixed electric dipole but different magnetic dipole Mie coefficients: $b_1=\mathrm{i}\frac{\gamma_z}{k_0}a_1$, $b_1=\mathrm{2i}\frac{\gamma_z}{k_0}a_1$ and $b_1=\mathrm{3i}\frac{\gamma_z}{k_0}a_1$, where $\gamma_z=0.142k_0$.} 
\label{fig:3}
\end{figure}
As a second example, we investigate the use of weakly evanescent standing waves as shown in Fig. \ref{fig:1}(c) for ultra-sensitive phase measurements. At a fixed location, the imaginary Poynting vector direction of an evanescent interference field with $k_{\parallel}=1.01k_0$ responds even more sensitively to phase changes. This time we will discuss sensitivity to phase changes in the incident plane waves. The principle is identical, because a relative phase difference of $\Delta\phi$ between the two interfering beams is equivalent to a displacement $y = \Delta\phi / 2k_{\parallel}$. One can derive from Eq. \ref{eq:6} that, for small phase changes, a sensitivity of 
\begin{align}
\frac{\mathrm{d}\theta_{\mathrm{s}}}{d\Delta\phi}|_{\Delta\phi\to0}=-\frac{k_{\parallel}}{2\gamma_z}, \label{eq:8}
\end{align}
can be reached. For a dipolar nanoparticle placed at the fixed location $y=0$ of the interference field with $k_{\parallel}/k_0=1.01$, a $24.25^{\circ}$ rotation of the maximum scattering direction in the upper $k$-space is introduced by a relative phase change of $\pi/50$ rad which is equivalent to a displacement of only $\lambda/101$ as shown in Fig. \ref{fig:3}(b,c). With smaller $\gamma_z/k_0$, even higher sensitivity can be expected. One might think that this comes at a price: if Kerker's condition were to be met with such excitation field at $y=0$ following Eq. \ref{eq:1}, the nanoparticle would need to have polarizabilities that fulfill $b_1=\mathrm{i}\frac{\gamma_z}{k_0}a_1$, resulting on a weak total scattering cross section proportional to $(\gamma_z/k_0)^2$. However, since in this case only the preferred scattering direction, aligned to the imaginary Poynting vector, is relevant for the detection in $k$-space, the Kerker's condition is not strictly required. As is shown in Fig. \ref{fig:3}(b) and (c), by using particles with larger $|b_1/a_1|$, the scattering power can be greatly improved without diminishing the sensitivity.

The proposed method exploits the full electromagnetic nature of the structured illumination formed by interferometric evanescent waves. By using a dipolar nano-antenna with both electric and magnetic responses, the fast-changing imaginary Poynting vector due to tiny displacement and phase variations is converted into sensitive scattering changes in $k$-space. We show that using the excitation field formed by strongly-evanescent standing waves, high sensitivity of the zero scattering direction with displacement can be observed over a long range. With weakly-evanescent wave interference, even higher sensitivity can be reached with tiny relative phase changes. The high sensitivity to displacement and phase changes, as demonstrated by this work, can form the basis for many applications including sensing, single molecule tracking, quantum metrology, wafer overlay and nano-optomechanical systems among many others.

\textit{Acknowledgement}: This work was supported by European Research Council Starting Grant ERC-2016-STG-714151-PSINFONI.


\begin{thebibliography}{36}%
\makeatletter
\providecommand \@ifxundefined [1]{%
 \@ifx{#1\undefined}
}%
\providecommand \@ifnum [1]{%
 \ifnum #1\expandafter \@firstoftwo
 \else \expandafter \@secondoftwo
 \fi
}%
\providecommand \@ifx [1]{%
 \ifx #1\expandafter \@firstoftwo
 \else \expandafter \@secondoftwo
 \fi
}%
\providecommand \natexlab [1]{#1}%
\providecommand \enquote  [1]{``#1''}%
\providecommand \bibnamefont  [1]{#1}%
\providecommand \bibfnamefont [1]{#1}%
\providecommand \citenamefont [1]{#1}%
\providecommand \href@noop [0]{\@secondoftwo}%
\providecommand \href [0]{\begingroup \@sanitize@url \@href}%
\providecommand \@href[1]{\@@startlink{#1}\@@href}%
\providecommand \@@href[1]{\endgroup#1\@@endlink}%
\providecommand \@sanitize@url [0]{\catcode `\\12\catcode `\$12\catcode
  `\&12\catcode `\#12\catcode `\^12\catcode `\_12\catcode `\%12\relax}%
\providecommand \@@startlink[1]{}%
\providecommand \@@endlink[0]{}%
\providecommand \url  [0]{\begingroup\@sanitize@url \@url }%
\providecommand \@url [1]{\endgroup\@href {#1}{\urlprefix }}%
\providecommand \urlprefix  [0]{URL }%
\providecommand \Eprint [0]{\href }%
\providecommand \doibase [0]{http://dx.doi.org/}%
\providecommand \selectlanguage [0]{\@gobble}%
\providecommand \bibinfo  [0]{\@secondoftwo}%
\providecommand \bibfield  [0]{\@secondoftwo}%
\providecommand \translation [1]{[#1]}%
\providecommand \BibitemOpen [0]{}%
\providecommand \bibitemStop [0]{}%
\providecommand \bibitemNoStop [0]{.\EOS\space}%
\providecommand \EOS [0]{\spacefactor3000\relax}%
\providecommand \BibitemShut  [1]{\csname bibitem#1\endcsname}%
\let\auto@bib@innerbib\@empty
\bibitem [{\citenamefont {Bobroff}(1993)}]{NBobroff1993}%
  \BibitemOpen
  \bibfield  {author} {\bibinfo {author} {\bibfnamefont {N.}~\bibnamefont
  {Bobroff}},\ }\href@noop {} {\bibfield  {journal} {\bibinfo  {journal} {Meas.
  Sci. Technol.}\ }\textbf {\bibinfo {volume} {4}},\ \bibinfo {pages} {907}
  (\bibinfo {year} {1993})}\BibitemShut {NoStop}%
\bibitem [{\citenamefont {Abott}\ \emph {et~al.}(2016)\citenamefont {Abott}
  \emph {et~al.}}]{BPAbbott2016}%
  \BibitemOpen
  \bibfield  {author} {\bibinfo {author} {\bibfnamefont {B.}~\bibnamefont
  {Abott}} \emph {et~al.},\ }\href@noop {} {\bibfield  {journal} {\bibinfo
  {journal} {Physical Review Letters}\ }\textbf {\bibinfo {volume} {116}},\
  \bibinfo {pages} {061102} (\bibinfo {year} {2016})}\BibitemShut {NoStop}%
\bibitem [{\citenamefont {Putman}\ \emph {et~al.}(1992)\citenamefont {Putman},
  \citenamefont {De~Grooth}, \citenamefont {Van~Hulst},\ and\ \citenamefont
  {Greve}}]{CAJPutman1992}%
  \BibitemOpen
  \bibfield  {author} {\bibinfo {author} {\bibfnamefont {C.~A.~J.}\
  \bibnamefont {Putman}}, \bibinfo {author} {\bibfnamefont {B.~G.}\
  \bibnamefont {De~Grooth}}, \bibinfo {author} {\bibfnamefont {N.~F.}\
  \bibnamefont {Van~Hulst}}, \ and\ \bibinfo {author} {\bibfnamefont
  {J.}~\bibnamefont {Greve}},\ }\href@noop {} {\bibfield  {journal} {\bibinfo
  {journal} {J. of Applied Physics}\ }\textbf {\bibinfo {volume} {72}},\
  \bibinfo {pages} {6} (\bibinfo {year} {1992})}\BibitemShut {NoStop}%
\bibitem [{\citenamefont {Barnett}\ \emph {et~al.}(2002)\citenamefont
  {Barnett}, \citenamefont {Fabre},\ and\ \citenamefont
  {Ma\^{i}tre}}]{SMBarnett2002}%
  \BibitemOpen
  \bibfield  {author} {\bibinfo {author} {\bibfnamefont {S.~M.}\ \bibnamefont
  {Barnett}}, \bibinfo {author} {\bibfnamefont {C.}~\bibnamefont {Fabre}}, \
  and\ \bibinfo {author} {\bibnamefont {Ma\^{i}tre}},\ }\href@noop {}
  {\bibfield  {journal} {\bibinfo  {journal} {Eur. Phys. J. D}\ }\textbf
  {\bibinfo {volume} {22}},\ \bibinfo {pages} {513} (\bibinfo {year}
  {2002})}\BibitemShut {NoStop}%
\bibitem [{\citenamefont {Gelles}\ \emph {et~al.}(1988)\citenamefont {Gelles},
  \citenamefont {Schnapp},\ and\ \citenamefont {Sheetz}}]{JGelles1988}%
  \BibitemOpen
  \bibfield  {author} {\bibinfo {author} {\bibfnamefont {J.}~\bibnamefont
  {Gelles}}, \bibinfo {author} {\bibfnamefont {B.~J.}\ \bibnamefont {Schnapp}},
  \ and\ \bibinfo {author} {\bibfnamefont {M.~P.}\ \bibnamefont {Sheetz}},\
  }\href@noop {} {\bibfield  {journal} {\bibinfo  {journal} {Nature}\ }\textbf
  {\bibinfo {volume} {331}},\ \bibinfo {pages} {450} (\bibinfo {year}
  {1988})}\BibitemShut {NoStop}%
\bibitem [{\citenamefont {Ortega~Arroyo}\ and\ \citenamefont
  {Kukura}(2015)}]{JOrtegaArroyo2015}%
  \BibitemOpen
  \bibfield  {author} {\bibinfo {author} {\bibfnamefont {J.}~\bibnamefont
  {Ortega~Arroyo}}\ and\ \bibinfo {author} {\bibfnamefont {P.}~\bibnamefont
  {Kukura}},\ }\href@noop {} {\bibfield  {journal} {\bibinfo  {journal} {Nature
  Photonics}\ }\textbf {\bibinfo {volume} {10}},\ \bibinfo {pages} {11}
  (\bibinfo {year} {2015})}\BibitemShut {NoStop}%
\bibitem [{\citenamefont {Balzarotti}\ \emph {et~al.}(2017)\citenamefont
  {Balzarotti}, \citenamefont {Eilers} \emph {et~al.}}]{FBalzarotti2017}%
  \BibitemOpen
  \bibfield  {author} {\bibinfo {author} {\bibfnamefont {F.}~\bibnamefont
  {Balzarotti}}, \bibinfo {author} {\bibfnamefont {Y.}~\bibnamefont {Eilers}},
  \emph {et~al.},\ }\href@noop {} {\bibfield  {journal} {\bibinfo  {journal}
  {Science}\ }\textbf {\bibinfo {volume} {355}},\ \bibinfo {pages} {601}
  (\bibinfo {year} {2017})}\BibitemShut {NoStop}%
\bibitem [{\citenamefont {Meyer}\ and\ \citenamefont
  {Amer}(1988)}]{GMeyer1988}%
  \BibitemOpen
  \bibfield  {author} {\bibinfo {author} {\bibfnamefont {G.}~\bibnamefont
  {Meyer}}\ and\ \bibinfo {author} {\bibfnamefont {N.~M.}\ \bibnamefont
  {Amer}},\ }\href@noop {} {\bibfield  {journal} {\bibinfo  {journal} {Appl.
  Phys. Lett.}\ }\textbf {\bibinfo {volume} {53}},\ \bibinfo {pages} {1045}
  (\bibinfo {year} {1988})}\BibitemShut {NoStop}%
\bibitem [{\citenamefont {Sanii}\ and\ \citenamefont
  {Ashby}(2010)}]{BSanii2010}%
  \BibitemOpen
  \bibfield  {author} {\bibinfo {author} {\bibfnamefont {B.}~\bibnamefont
  {Sanii}}\ and\ \bibinfo {author} {\bibfnamefont {P.~D.}\ \bibnamefont
  {Ashby}},\ }\href@noop {} {\bibfield  {journal} {\bibinfo  {journal}
  {Physical Review Letters}\ }\textbf {\bibinfo {volume} {104}},\ \bibinfo
  {pages} {147203} (\bibinfo {year} {2010})}\BibitemShut {NoStop}%
\bibitem [{\citenamefont {Gloppe}\ \emph {et~al.}(2014)\citenamefont {Gloppe},
  \citenamefont {Verlot}, \citenamefont {Dupont-Ferrier} \emph
  {et~al.}}]{AGloppe2014}%
  \BibitemOpen
  \bibfield  {author} {\bibinfo {author} {\bibfnamefont {A.}~\bibnamefont
  {Gloppe}}, \bibinfo {author} {\bibfnamefont {P.}~\bibnamefont {Verlot}},
  \bibinfo {author} {\bibfnamefont {E.}~\bibnamefont {Dupont-Ferrier}},  \emph
  {et~al.},\ }\href@noop {} {\bibfield  {journal} {\bibinfo  {journal} {Nature
  Nanotech.}\ }\textbf {\bibinfo {volume} {9}},\ \bibinfo {pages} {920}
  (\bibinfo {year} {2014})}\BibitemShut {NoStop}%
\bibitem [{\citenamefont {Bl\={u}ms}\ \emph {et~al.}(2018)\citenamefont
  {Bl\={u}ms}, \citenamefont {Piotrowski}, \citenamefont {Hussain} \emph
  {et~al.}}]{VBlums2018}%
  \BibitemOpen
  \bibfield  {author} {\bibinfo {author} {\bibnamefont {Bl\={u}ms}}, \bibinfo
  {author} {\bibfnamefont {M.}~\bibnamefont {Piotrowski}}, \bibinfo {author}
  {\bibfnamefont {M.~I.}\ \bibnamefont {Hussain}},  \emph {et~al.},\
  }\href@noop {} {\bibfield  {journal} {\bibinfo  {journal} {Sci. Adv.}\
  }\textbf {\bibinfo {volume} {4}},\ \bibinfo {pages} {eaao4453} (\bibinfo
  {year} {2018})}\BibitemShut {NoStop}%
\bibitem [{\citenamefont {Nugent-Glandorf}\ and\ \citenamefont
  {Perkins}(2014)}]{LNugentGlandorf2004}%
  \BibitemOpen
  \bibfield  {author} {\bibinfo {author} {\bibfnamefont {L.}~\bibnamefont
  {Nugent-Glandorf}}\ and\ \bibinfo {author} {\bibnamefont {Perkins}},\
  }\href@noop {} {\bibfield  {journal} {\bibinfo  {journal} {Optics Letters}\
  }\textbf {\bibinfo {volume} {29}},\ \bibinfo {pages} {2611} (\bibinfo {year}
  {2014})}\BibitemShut {NoStop}%
\bibitem [{\citenamefont {den Boef}(2016)}]{AJdenBoef2016}%
  \BibitemOpen
  \bibfield  {author} {\bibinfo {author} {\bibfnamefont {A.~J.}\ \bibnamefont
  {den Boef}},\ }\href@noop {} {\bibfield  {journal} {\bibinfo  {journal}
  {Surf. Topogr.: Metrol. Prop.}\ }\textbf {\bibinfo {volume} {4}},\ \bibinfo
  {pages} {023001} (\bibinfo {year} {2016})}\BibitemShut {NoStop}%
\bibitem [{\citenamefont {Treps}\ \emph {et~al.}(2002)\citenamefont {Treps},
  \citenamefont {Andersen}, \citenamefont {Buchler} \emph
  {et~al.}}]{NTreps2002}%
  \BibitemOpen
  \bibfield  {author} {\bibinfo {author} {\bibfnamefont {N.}~\bibnamefont
  {Treps}}, \bibinfo {author} {\bibfnamefont {U.}~\bibnamefont {Andersen}},
  \bibinfo {author} {\bibfnamefont {B.}~\bibnamefont {Buchler}},  \emph
  {et~al.},\ }\href@noop {} {\bibfield  {journal} {\bibinfo  {journal}
  {Physical Review Letters}\ }\textbf {\bibinfo {volume} {88}},\ \bibinfo
  {pages} {203601} (\bibinfo {year} {2002})}\BibitemShut {NoStop}%
\bibitem [{\citenamefont {Pooser}\ and\ \citenamefont
  {Lawrie}(2015)}]{RCPooser2015}%
  \BibitemOpen
  \bibfield  {author} {\bibinfo {author} {\bibfnamefont {R.~C.}\ \bibnamefont
  {Pooser}}\ and\ \bibinfo {author} {\bibfnamefont {B.}~\bibnamefont
  {Lawrie}},\ }\href@noop {} {\bibfield  {journal} {\bibinfo  {journal}
  {Optica}\ }\textbf {\bibinfo {volume} {2}},\ \bibinfo {pages} {393} (\bibinfo
  {year} {2015})}\BibitemShut {NoStop}%
\bibitem [{\citenamefont {Ben~Dixon}\ \emph {et~al.}(2009)\citenamefont
  {Ben~Dixon}, \citenamefont {Starling}, \citenamefont {Jordan},\ and\
  \citenamefont {Howell}}]{PBenDixon2009}%
  \BibitemOpen
  \bibfield  {author} {\bibinfo {author} {\bibfnamefont {P.}~\bibnamefont
  {Ben~Dixon}}, \bibinfo {author} {\bibfnamefont {D.~J.}\ \bibnamefont
  {Starling}}, \bibinfo {author} {\bibfnamefont {A.~N.}\ \bibnamefont
  {Jordan}}, \ and\ \bibinfo {author} {\bibfnamefont {J.~C.}\ \bibnamefont
  {Howell}},\ }\href@noop {} {\bibfield  {journal} {\bibinfo  {journal}
  {Physical Review Letters}\ }\textbf {\bibinfo {volume} {102}},\ \bibinfo
  {pages} {173601} (\bibinfo {year} {2009})}\BibitemShut {NoStop}%
\bibitem [{\citenamefont {Turner}\ \emph {et~al.}(2011)\citenamefont {Turner},
  \citenamefont {Charles}, \citenamefont {Schlamminger},\ and\ \citenamefont
  {Gundlach}}]{MDTurner2011}%
  \BibitemOpen
  \bibfield  {author} {\bibinfo {author} {\bibfnamefont {M.~D.}\ \bibnamefont
  {Turner}}, \bibinfo {author} {\bibfnamefont {A.~H.}\ \bibnamefont {Charles}},
  \bibinfo {author} {\bibfnamefont {S.}~\bibnamefont {Schlamminger}}, \ and\
  \bibinfo {author} {\bibfnamefont {J.~H.}\ \bibnamefont {Gundlach}},\
  }\href@noop {} {\bibfield  {journal} {\bibinfo  {journal} {Optics Letters}\
  }\textbf {\bibinfo {volume} {36}},\ \bibinfo {pages} {1479} (\bibinfo {year}
  {2011})}\BibitemShut {NoStop}%
\bibitem [{\citenamefont {Neugebauer}\ \emph {et~al.}(2016)\citenamefont
  {Neugebauer}, \citenamefont {Wo\'{z}niak}, \citenamefont {Bag}, \citenamefont
  {Leuchs},\ and\ \citenamefont {Banzer}}]{MNeugebauer2016}%
  \BibitemOpen
  \bibfield  {author} {\bibinfo {author} {\bibfnamefont {M.}~\bibnamefont
  {Neugebauer}}, \bibinfo {author} {\bibfnamefont {P.}~\bibnamefont
  {Wo\'{z}niak}}, \bibinfo {author} {\bibfnamefont {A.}~\bibnamefont {Bag}},
  \bibinfo {author} {\bibfnamefont {G.}~\bibnamefont {Leuchs}}, \ and\ \bibinfo
  {author} {\bibfnamefont {P.}~\bibnamefont {Banzer}},\ }\href@noop {}
  {\bibfield  {journal} {\bibinfo  {journal} {Nature, Comm.}\ }\textbf
  {\bibinfo {volume} {7}},\ \bibinfo {pages} {11286} (\bibinfo {year}
  {2016})}\BibitemShut {NoStop}%
\bibitem [{\citenamefont {Xi}\ \emph {et~al.}(2016{\natexlab{a}})\citenamefont
  {Xi}, \citenamefont {Wei}, \citenamefont {Adam},\ and\ \citenamefont
  {Urbach}}]{ZXi2016}%
  \BibitemOpen
  \bibfield  {author} {\bibinfo {author} {\bibfnamefont {Z.}~\bibnamefont
  {Xi}}, \bibinfo {author} {\bibfnamefont {L.}~\bibnamefont {Wei}}, \bibinfo
  {author} {\bibfnamefont {A.~J.~L.}\ \bibnamefont {Adam}}, \ and\ \bibinfo
  {author} {\bibfnamefont {H.~P.}\ \bibnamefont {Urbach}},\ }\href@noop {}
  {\bibfield  {journal} {\bibinfo  {journal} {Physical Review Letters}\
  }\textbf {\bibinfo {volume} {117}},\ \bibinfo {pages} {113903} (\bibinfo
  {year} {2016}{\natexlab{a}})}\BibitemShut {NoStop}%
\bibitem [{\citenamefont {Xi}\ and\ \citenamefont {Urbach}(2017)}]{ZXi2017}%
  \BibitemOpen
  \bibfield  {author} {\bibinfo {author} {\bibfnamefont {Z.}~\bibnamefont
  {Xi}}\ and\ \bibinfo {author} {\bibfnamefont {H.~P.}\ \bibnamefont
  {Urbach}},\ }\href@noop {} {\bibfield  {journal} {\bibinfo  {journal}
  {Physical Review Letters}\ }\textbf {\bibinfo {volume} {119}},\ \bibinfo
  {pages} {053902} (\bibinfo {year} {2017})}\BibitemShut {NoStop}%
\bibitem [{\citenamefont {Bag}\ \emph {et~al.}()\citenamefont {Bag},
  \citenamefont {Neugebauer}, \citenamefont {Wo\'{z}niak}, \citenamefont
  {Leuchs},\ and\ \citenamefont {Banzer}}]{ABag2018}%
  \BibitemOpen
  \bibfield  {author} {\bibinfo {author} {\bibfnamefont {A.}~\bibnamefont
  {Bag}}, \bibinfo {author} {\bibfnamefont {M.}~\bibnamefont {Neugebauer}},
  \bibinfo {author} {\bibfnamefont {P.}~\bibnamefont {Wo\'{z}niak}}, \bibinfo
  {author} {\bibfnamefont {G.}~\bibnamefont {Leuchs}}, \ and\ \bibinfo {author}
  {\bibfnamefont {P.}~\bibnamefont {Banzer}},\ }\href@noop {} {\ }\Eprint
  {http://arxiv.org/abs/1804.10176} {arXiv:1804.10176} \BibitemShut {NoStop}%
\bibitem [{\citenamefont {Kuznetsov}\ \emph {et~al.}(2016)\citenamefont
  {Kuznetsov}, \citenamefont {Miroshnichenko} \emph
  {et~al.}}]{AIKuznetsov2016}%
  \BibitemOpen
  \bibfield  {author} {\bibinfo {author} {\bibfnamefont {A.~I.}\ \bibnamefont
  {Kuznetsov}}, \bibinfo {author} {\bibfnamefont {A.~E.}\ \bibnamefont
  {Miroshnichenko}},  \emph {et~al.},\ }\href@noop {} {\bibfield  {journal}
  {\bibinfo  {journal} {Science}\ }\textbf {\bibinfo {volume} {354}},\ \bibinfo
  {pages} {aag2472} (\bibinfo {year} {2016})}\BibitemShut {NoStop}%
\bibitem [{\citenamefont {Wo\'{z}niak}\ \emph {et~al.}(2015)\citenamefont
  {Wo\'{z}niak}, \citenamefont {Banzer},\ and\ \citenamefont
  {Leuchs}}]{PWozniak2015}%
  \BibitemOpen
  \bibfield  {author} {\bibinfo {author} {\bibfnamefont {P.}~\bibnamefont
  {Wo\'{z}niak}}, \bibinfo {author} {\bibfnamefont {P.}~\bibnamefont {Banzer}},
  \ and\ \bibinfo {author} {\bibfnamefont {G.}~\bibnamefont {Leuchs}},\
  }\href@noop {} {\bibfield  {journal} {\bibinfo  {journal} {Laser Photonics
  Rev.}\ }\textbf {\bibinfo {volume} {9}},\ \bibinfo {pages} {231} (\bibinfo
  {year} {2015})}\BibitemShut {NoStop}%
\bibitem [{\citenamefont {Das}\ \emph {et~al.}(2015)\citenamefont {Das},
  \citenamefont {Iyer}, \citenamefont {DeCrescent},\ and\ \citenamefont
  {Schuller}}]{TDas2015}%
  \BibitemOpen
  \bibfield  {author} {\bibinfo {author} {\bibfnamefont {T.}~\bibnamefont
  {Das}}, \bibinfo {author} {\bibfnamefont {P.~P.}\ \bibnamefont {Iyer}},
  \bibinfo {author} {\bibfnamefont {R.~A.}\ \bibnamefont {DeCrescent}}, \ and\
  \bibinfo {author} {\bibfnamefont {J.~A.}\ \bibnamefont {Schuller}},\
  }\href@noop {} {\bibfield  {journal} {\bibinfo  {journal} {Phys. Rev. B}\
  }\textbf {\bibinfo {volume} {92}},\ \bibinfo {pages} {241110(R)} (\bibinfo
  {year} {2015})}\BibitemShut {NoStop}%
\bibitem [{\citenamefont {Xi}\ \emph {et~al.}(2016{\natexlab{b}})\citenamefont
  {Xi}, \citenamefont {Wei}, \citenamefont {Adam},\ and\ \citenamefont
  {Urbach}}]{ZXiOL2016}%
  \BibitemOpen
  \bibfield  {author} {\bibinfo {author} {\bibfnamefont {Z.}~\bibnamefont
  {Xi}}, \bibinfo {author} {\bibfnamefont {L.}~\bibnamefont {Wei}}, \bibinfo
  {author} {\bibfnamefont {A.~J.~L.}\ \bibnamefont {Adam}}, \ and\ \bibinfo
  {author} {\bibfnamefont {H.~P.}\ \bibnamefont {Urbach}},\ }\href@noop {}
  {\bibfield  {journal} {\bibinfo  {journal} {Optics Letters}\ }\textbf
  {\bibinfo {volume} {41}},\ \bibinfo {pages} {33} (\bibinfo {year}
  {2016}{\natexlab{b}})}\BibitemShut {NoStop}%
\bibitem [{\citenamefont {Wei}\ \emph {et~al.}(2017)\citenamefont {Wei},
  \citenamefont {Bhattacharya},\ and\ \citenamefont {Urbach}}]{LWEI2017}%
  \BibitemOpen
  \bibfield  {author} {\bibinfo {author} {\bibfnamefont {L.}~\bibnamefont
  {Wei}}, \bibinfo {author} {\bibfnamefont {N.}~\bibnamefont {Bhattacharya}}, \
  and\ \bibinfo {author} {\bibfnamefont {H.~P.}\ \bibnamefont {Urbach}},\
  }\href@noop {} {\bibfield  {journal} {\bibinfo  {journal} {Optics Letters}\
  }\textbf {\bibinfo {volume} {42}},\ \bibinfo {pages} {1776} (\bibinfo {year}
  {2017})}\BibitemShut {NoStop}%
\bibitem [{\citenamefont {Liu}\ and\ \citenamefont {Kivshar}(2018)}]{WLiu2018}%
  \BibitemOpen
  \bibfield  {author} {\bibinfo {author} {\bibfnamefont {W.}~\bibnamefont
  {Liu}}\ and\ \bibinfo {author} {\bibfnamefont {Y.~S.}\ \bibnamefont
  {Kivshar}},\ }\href@noop {} {\bibfield  {journal} {\bibinfo  {journal}
  {Optics Express}\ }\textbf {\bibinfo {volume} {26}},\ \bibinfo {pages}
  {13085} (\bibinfo {year} {2018})}\BibitemShut {NoStop}%
\bibitem [{\citenamefont {Bliokh}\ \emph {et~al.}(2015)\citenamefont {Bliokh},
  \citenamefont {Smirnova},\ and\ \citenamefont {Nori}}]{KYBliokh2015}%
  \BibitemOpen
  \bibfield  {author} {\bibinfo {author} {\bibfnamefont {K.~Y.}\ \bibnamefont
  {Bliokh}}, \bibinfo {author} {\bibfnamefont {D.}~\bibnamefont {Smirnova}}, \
  and\ \bibinfo {author} {\bibfnamefont {F.}~\bibnamefont {Nori}},\ }\href@noop
  {} {\bibfield  {journal} {\bibinfo  {journal} {Science}\ }\textbf {\bibinfo
  {volume} {348}},\ \bibinfo {pages} {1448} (\bibinfo {year}
  {2015})}\BibitemShut {NoStop}%
\bibitem [{\citenamefont {Wei}\ \emph {et~al.}()\citenamefont {Wei},
  \citenamefont {Picardi}, \citenamefont {Kingsley-Smith}, \citenamefont
  {Zayats},\ and\ \citenamefont {Rodr\'{i}guez-Fortu\~{n}o}}]{LWei2018}%
  \BibitemOpen
  \bibfield  {author} {\bibinfo {author} {\bibfnamefont {L.}~\bibnamefont
  {Wei}}, \bibinfo {author} {\bibfnamefont {M.~F.}\ \bibnamefont {Picardi}},
  \bibinfo {author} {\bibfnamefont {J.~J.}\ \bibnamefont {Kingsley-Smith}},
  \bibinfo {author} {\bibfnamefont {A.~V.}\ \bibnamefont {Zayats}}, \ and\
  \bibinfo {author} {\bibfnamefont {F.~J.}\ \bibnamefont
  {Rodr\'{i}guez-Fortu\~{n}o}},\ }\href@noop {} {\ }\Eprint
  {http://arxiv.org/abs/1803.04821} {arXiv:1803.04821} \BibitemShut {NoStop}%
\bibitem [{\citenamefont {Xiang}\ \emph {et~al.}(2018)\citenamefont {Xiang},
  \citenamefont {Li} \emph {et~al.}}]{JXiang2018}%
  \BibitemOpen
  \bibfield  {author} {\bibinfo {author} {\bibfnamefont {J.}~\bibnamefont
  {Xiang}}, \bibinfo {author} {\bibfnamefont {J.}~\bibnamefont {Li}},  \emph
  {et~al.},\ }\href@noop {} {\bibfield  {journal} {\bibinfo  {journal} {Laser
  Photonics Rev.}\ }\textbf {\bibinfo {volume} {2018}},\ \bibinfo {pages}
  {1800032} (\bibinfo {year} {2018})}\BibitemShut {NoStop}%
\bibitem [{\citenamefont {Nieto-Vesperinas}\ and\ \citenamefont
  {Saenz}(2010)}]{MNietoVesperinas2010}%
  \BibitemOpen
  \bibfield  {author} {\bibinfo {author} {\bibfnamefont {M.}~\bibnamefont
  {Nieto-Vesperinas}}\ and\ \bibinfo {author} {\bibfnamefont {J.~J.}\
  \bibnamefont {Saenz}},\ }\href@noop {} {\bibfield  {journal} {\bibinfo
  {journal} {Optics Letters}\ }\textbf {\bibinfo {volume} {35}},\ \bibinfo
  {pages} {4078} (\bibinfo {year} {2010})}\BibitemShut {NoStop}%
\bibitem [{\citenamefont {Bliokh}\ \emph {et~al.}(2014)\citenamefont {Bliokh},
  \citenamefont {Bekshaev},\ and\ \citenamefont {Nori}}]{KYBliokh2014}%
  \BibitemOpen
  \bibfield  {author} {\bibinfo {author} {\bibfnamefont {K.~Y.}\ \bibnamefont
  {Bliokh}}, \bibinfo {author} {\bibfnamefont {A.~Y.}\ \bibnamefont
  {Bekshaev}}, \ and\ \bibinfo {author} {\bibfnamefont {F.}~\bibnamefont
  {Nori}},\ }\href@noop {} {\bibfield  {journal} {\bibinfo  {journal} {Nature
  Comm.}\ }\textbf {\bibinfo {volume} {5}},\ \bibinfo {pages} {3300} (\bibinfo
  {year} {2014})}\BibitemShut {NoStop}%
\bibitem [{\citenamefont {Li}\ \emph {et~al.}(2015)\citenamefont {Li},
  \citenamefont {Shao} \emph {et~al.}}]{DLi2015}%
  \BibitemOpen
  \bibfield  {author} {\bibinfo {author} {\bibfnamefont {D.}~\bibnamefont
  {Li}}, \bibinfo {author} {\bibfnamefont {L.}~\bibnamefont {Shao}},  \emph
  {et~al.},\ }\href@noop {} {\bibfield  {journal} {\bibinfo  {journal}
  {Science}\ }\textbf {\bibinfo {volume} {349}},\ \bibinfo {pages} {6251}
  (\bibinfo {year} {2015})}\BibitemShut {NoStop}%
\bibitem [{\citenamefont {Nieto-Vesperinas}\ \emph {et~al.}(2010)\citenamefont
  {Nieto-Vesperinas}, \citenamefont {Saenz}, \citenamefont {G\'{o}mez-Medina},\
  and\ \citenamefont {Chantada}}]{MNietoVesperinasOE2010}%
  \BibitemOpen
  \bibfield  {author} {\bibinfo {author} {\bibfnamefont {M.}~\bibnamefont
  {Nieto-Vesperinas}}, \bibinfo {author} {\bibfnamefont {J.~J.}\ \bibnamefont
  {Saenz}}, \bibinfo {author} {\bibfnamefont {R.}~\bibnamefont
  {G\'{o}mez-Medina}}, \ and\ \bibinfo {author} {\bibfnamefont
  {L.}~\bibnamefont {Chantada}},\ }\href@noop {} {\bibfield  {journal}
  {\bibinfo  {journal} {Optics Express}\ }\textbf {\bibinfo {volume} {18}},\
  \bibinfo {pages} {11428} (\bibinfo {year} {2010})}\BibitemShut {NoStop}%
\bibitem [{\citenamefont {Fu}\ \emph {et~al.}(2013)\citenamefont {Fu},
  \citenamefont {Kuznetsov}, \citenamefont {Miroshnichenko}, \citenamefont
  {Yu},\ and\ \citenamefont {Luk'yanchuk}}]{YHFu2013}%
  \BibitemOpen
  \bibfield  {author} {\bibinfo {author} {\bibfnamefont {Y.~H.}\ \bibnamefont
  {Fu}}, \bibinfo {author} {\bibfnamefont {A.~I.}\ \bibnamefont {Kuznetsov}},
  \bibinfo {author} {\bibfnamefont {A.~E.}\ \bibnamefont {Miroshnichenko}},
  \bibinfo {author} {\bibfnamefont {Y.~F.}\ \bibnamefont {Yu}}, \ and\ \bibinfo
  {author} {\bibfnamefont {B.}~\bibnamefont {Luk'yanchuk}},\ }\href@noop {}
  {\bibfield  {journal} {\bibinfo  {journal} {Nature, Comm.}\ }\textbf
  {\bibinfo {volume} {4}},\ \bibinfo {pages} {1527} (\bibinfo {year}
  {2013})}\BibitemShut {NoStop}%
\bibitem [{\citenamefont {Picardi}\ \emph {et~al.}(2017)\citenamefont
  {Picardi}, \citenamefont {Manjavacas}, \citenamefont {Zayats},\ and\
  \citenamefont {Rodr\'{i}guez-Fortu\~{n}o}}]{MFPicardi2017}%
  \BibitemOpen
  \bibfield  {author} {\bibinfo {author} {\bibfnamefont {M.~F.}\ \bibnamefont
  {Picardi}}, \bibinfo {author} {\bibfnamefont {A.}~\bibnamefont {Manjavacas}},
  \bibinfo {author} {\bibfnamefont {A.~V.}\ \bibnamefont {Zayats}}, \ and\
  \bibinfo {author} {\bibfnamefont {F.~J.}\ \bibnamefont
  {Rodr\'{i}guez-Fortu\~{n}o}},\ }\href@noop {} {\bibfield  {journal} {\bibinfo
   {journal} {Phys. Rev. B}\ }\textbf {\bibinfo {volume} {95}},\ \bibinfo
  {pages} {245416} (\bibinfo {year} {2017})}\BibitemShut {NoStop}%
\end{thebibliography}
\end{document}